\documentclass[aps,preprint,prl,superscriptaddress]{revtex4}

\usepackage{graphicx}
\usepackage{amsmath}
\usepackage{amssymb}
\usepackage{units}
\usepackage{color}

\usepackage{setspace}

\newcommand{\etal}{{\it et al.}\,}
\begin{document}

\title{Population Inversion in Monolayer and Bilayer Graphene}

\author{Isabella Gierz}
\email{Isabella.Gierz@mpsd.mpg.de}
\affiliation{Max Planck Institute for the Structure and Dynamics of Matter, Hamburg, Germany}
\author{Matteo Mitrano}
\affiliation{Max Planck Institute for the Structure and Dynamics of Matter, Hamburg, Germany}
\author{Jesse C. Petersen}
\affiliation{Department of Physics, Clarendon Laboratory, University of Oxford, Oxford, United Kingdom}
\author{Cephise Cacho}
\author{I. C. Edmond Turcu}
\author{Emma Springate}
\affiliation{Central Laser Facility, STFC Rutherford Appleton Laboratory, Harwell, United Kingdom}
\author{Alexander St\"ohr}
\author{Axel K\"ohler}
\author{Ulrich Starke}
\affiliation{Max Planck Institute for Solid State Research, Stuttgart, Germany}
\author{Andrea Cavalleri}
\affiliation{Max Planck Institute for the Structure and Dynamics of Matter, Hamburg, Germany}
\affiliation{Department of Physics, Clarendon Laboratory, University of Oxford, Oxford, United Kingdom}

\date{\today}

\maketitle

{\bf The recent demonstration of saturable absorption and negative optical conductivity in the Terahertz range in graphene has opened up new opportunities for optoelectronic applications based on this and other low dimensional materials. Recently, population inversion across the Dirac point has been observed directly by time- and angle-resolved photoemission spectroscopy (tr-ARPES), revealing a relaxation time of only $\sim130$ femtoseconds. This severely limits the applicability of single layer graphene to, for example, Terahertz light amplification. Here we use tr-ARPES to demonstrate long-lived population inversion in bilayer graphene. The effect is attributed to the small band gap found in this compound. We propose a microscopic model for these observations and speculate that an enhancement of both the pump photon energy and the pump fluence may further increase this lifetime.}

\pagebreak

\section{Introduction}

The high charge-carrier mobility in graphene has raised hopes for future high-speed electronic devices. However, the interaction of hot carriers with the lattice may limit the operation of such devices. Thus, significant efforts were recently dedicated to understand the cooling dynamics of hot carriers by time-resolved optical spectroscopy. More recently, new phenomena like saturable absorption, population inversion, and possibly carrier multiplication, were discussed, broadening the scope of dynamical studies in graphene. In the meantime, graphene has been used successfully to generate ultrashort laser pulses \cite{Bao_2009,Sun_2010}. Applications as an amplifying medium in a Terahertz laser may be feasible \cite{Rhyzhii_2007,Li_2012,Winzer_2013,Gierz_2013}. 

Direct access to the time evolution of photoexcited carriers is made possible by time- and angle-resolved photoemission spectroscopy (tr-ARPES), a technique that measures the number of electrons as a function of energy, momentum, and time. Here we focus on the stage immediately after photoexcitation with a femtosecond laser pulse, where the carriers in the conduction band are decoupled from the carriers in the valence band by a bottleneck at the Dirac point. Two electron gases with distinct chemical potentials and temperatures are then observed. We measure the formation of this population inverted state after high-fluence interband excitation and compare its lifetime in monolayer graphene (a zero-gap semimetal with linearly dispersing bands) and bilayer graphene (a narrow-gap semiconductor with conventional parabolic dispersion) finding longer lifetimes in the latter. We attribute this observation to a stronger relaxation bottleneck in the presence of a band gap.

\section{Methods}

\subsection{Time- and Angle-Resolved Photoemission Spectroscopy}

Femtosecond laser pulses tuneable between 800\,nm and 15\,$\mu$m wavelength are generally used in our apparatus to excite the sample. At a variable delay from this excitation, the sample is illuminated with a second extreme ultra-violet (EUV) pulse. Photoelectrons are measured with a hemispherical analyzer yielding the photocurrent as a function of kinetic energy, $E_{\text{kin}}$, and emission angle, $\theta$. From the kinetic energy, the binding energy, $E_{\text{B}}$, of the electrons in the solid can be obtained as $E_{\text{B}}=\hbar\omega-\phi-E_{\text{kin}}$, where $\hbar\omega$ is the EUV photon energy and $\phi$ is the work function of the analyzer. The measured emission angle can be converted into in-plane momentum $k_{||}=0.512\sin\theta\sqrt{E_{\text{kin}}}$, so that the image on the two-dimensional detector directly maps the band structure along a particular cut in momentum space.

Pump and probe pulses are generated by parametric downconversion and high-order harmonic upconversion, respectively, of 30\,fs - 780\,nm - 1\,kHz pulses from a Ti:Sapphire laser. The pump pulses are generated with a commercial optical parametric amplifier (1 to 2\,$\mu$m) with subsequent difference frequency generation (4 to 15\,$\mu$m). About 1\,mJ of laser energy is used for high-order harmonics generation in an argon gas jet, generating EUV photons in the energy range between 20 and 40\,eV. A single-grating time-preserving monochromator is used to select a single harmonic at $\hbar\omega_{\text{probe}}=31.5$\,eV, yielding $<30$\,fs long EUV pulses with a spectral width of about 300\,meV \cite{Frassetto_2011}. The EUV probe is focused onto the sample with a toroidal mirror resulting in a typical spot size of 200\,$\mu$m.

\subsection{Sample Preparation}

\begin{figure}
	\center
  \includegraphics[width = 0.7\columnwidth]{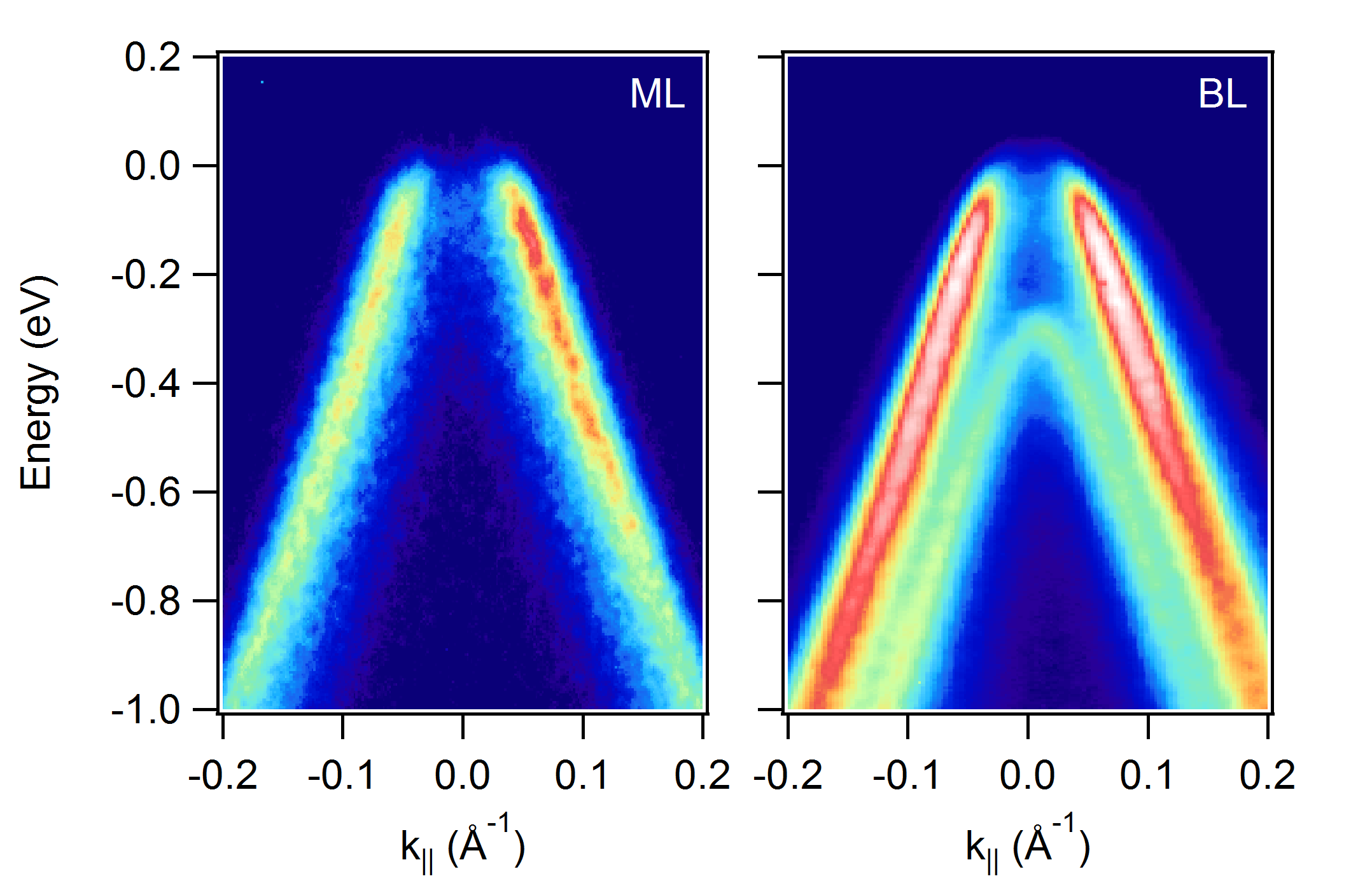}
  \caption{Equilibrium band structure of a hydrogen-intercalated graphene monolayer (ML) and bilayer (BL) measured with He II $\alpha$ radiation.}
  \label{fig_static}
\end{figure}

Epitaxial monolayer and bilayer graphene samples for the present investigation were grown at the Max Planck Institute for Solid State Research. In a first step, the SiC(0001) substrates were hydrogen-etched to obtain large flat terraces. In a second step, the substrates were annealed in Ar atmosphere to graphitize the surface, resulting in a carbon coverage of one and two monolayers, respectively. The last preparation step ensured that the carbon layers were decoupled from the underlying substrate by hydrogen intercalation, resulting in quasi-freestanding monolayer (ML) and bilayer (BL) graphene \cite{Riedl_2009}. The static band structure of these samples, measured along a cut through the K-point perpendicular to the $\Gamma$K-direction, is shown in Fig. \ref{fig_static}. In both cases the top of the valence band is unoccupied, indicating a slight hole doping with the chemical potential at about $-200$\,meV. The bilayer sample has a small band gap of about 200\,meV due to the different amounts of charges in the top and bottom layer, respectively, induced by charge transfer from the substrate \cite{Ulstrup_2014}. The samples were exposed to air and reinserted into ultra-high vacuum at the Artemis user facility in Harwell, United Kingdom, where the tr-ARPES experiments were performed. After a mild annealing at $\sim200^{\circ}$C the original band structure was recovered.

\section{Results}

\begin{figure}
	\center
  \includegraphics[width = 1\columnwidth]{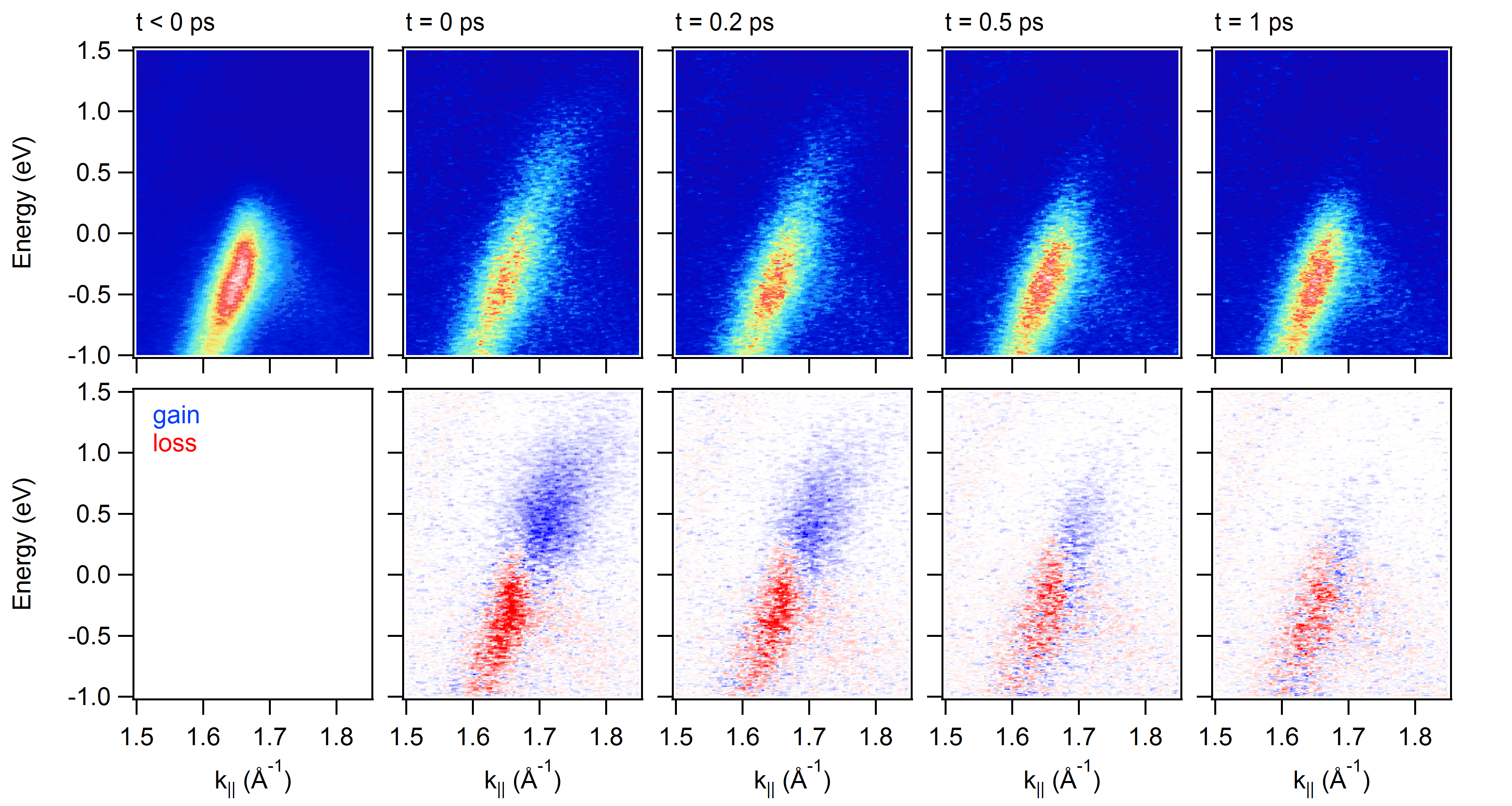}
  \caption{Upper panel: snapshots of the electronic structure of monolayer graphene in the vicinity of the K-point around the Fermi level for different pump-probe delays. Lower panel: corresponding pump-induced changes of the photocurrent. This data set is the same as in Ref. \cite{Gierz_2013}.}
  \label{fig_snapshots_ML}
\end{figure}

\begin{figure}
	\center
  \includegraphics[width = 1\columnwidth]{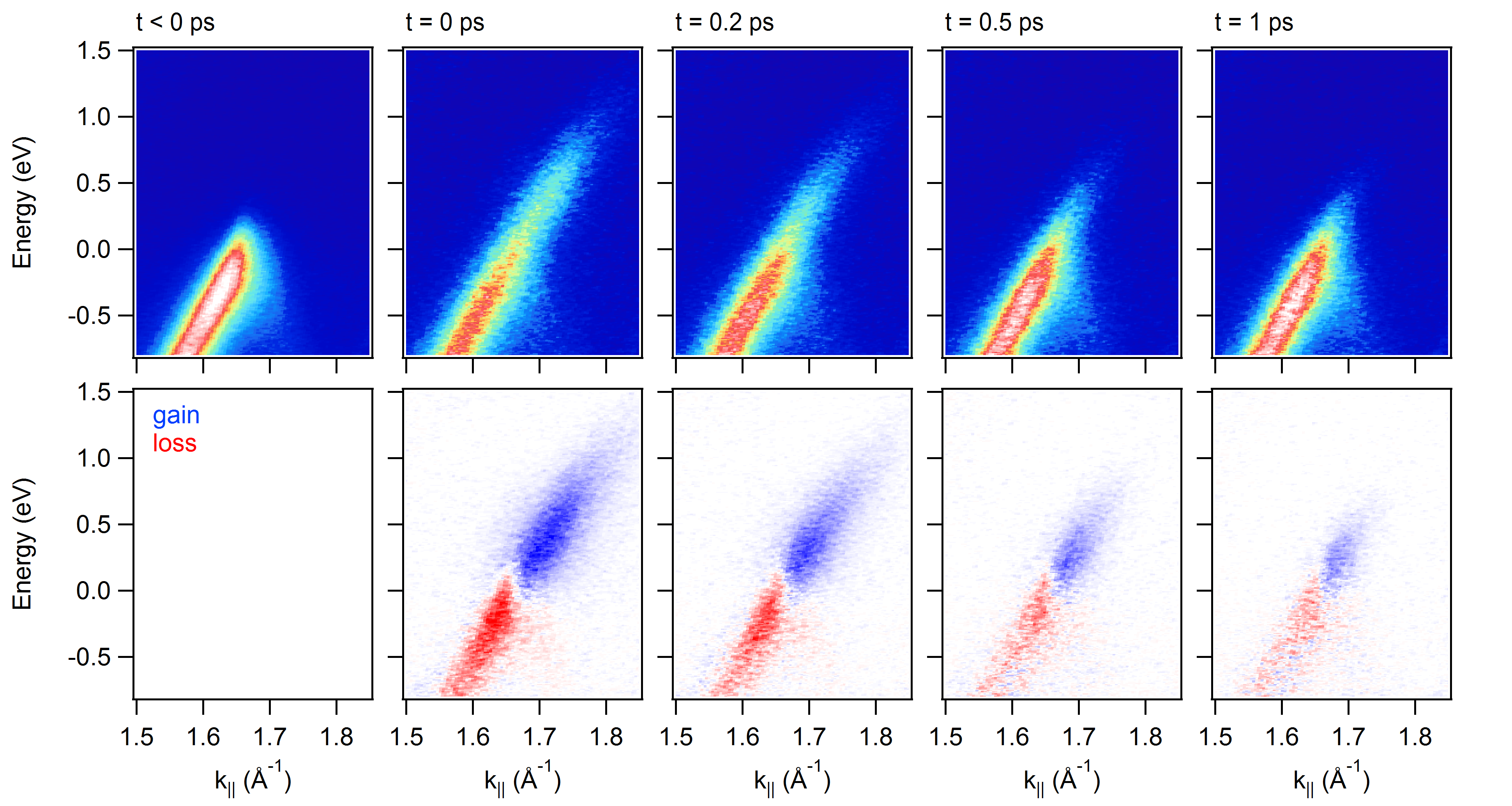}
  \caption{Upper panel: snapshots of the electronic structure of bilayer graphene in the vicinity of the K-point around the Fermi level for different pump-probe delays. Lower panel: corresponding pump-induced changes of the photocurrent.}
  \label{fig_snapshots_BL}
\end{figure}

Figures \ref{fig_snapshots_ML} and \ref{fig_snapshots_BL} show snapshots of the band structure through the K-point along the $\Gamma$K-direction for different pump-probe delays (upper panel) along with the pump-induced changes of the photocurrent (lower panel) for monolayer and bilayer graphene, respectively. The monolayer sample was excited using $\hbar\omega_{\text{pump}}=1.3$\,$\mu$m at a fluence of $F=4.6$\,mJ/cm$^2$, the bilayer sample was excited using $\hbar\omega_{\text{pump}}=1.5$\,$\mu$m at a fluence of $F=4.1$\,mJ/cm$^2$. For cuts along the $\Gamma$K-direction, only one branch of the $\pi$-bands is visible due to photoemission matrix element effects \cite{Shirley_1995}. Due to the limited energy resolution of the pulsed EUV source, the two $\pi$-bands of the bilayer cannot be distinguished in the time-resolved data. The data set for the graphene monolayer is the same as in Ref. \cite{Gierz_2013}.

Upon arrival of the pump pulse the intensity of the $\pi$-bands below the Fermi level was observed to reduce, and carriers were injected in the previously unoccupied states. Within about one picosecond the excited carriers relax back to their equilibrium position.

\begin{figure}
	\center
  \includegraphics[width = 1\columnwidth]{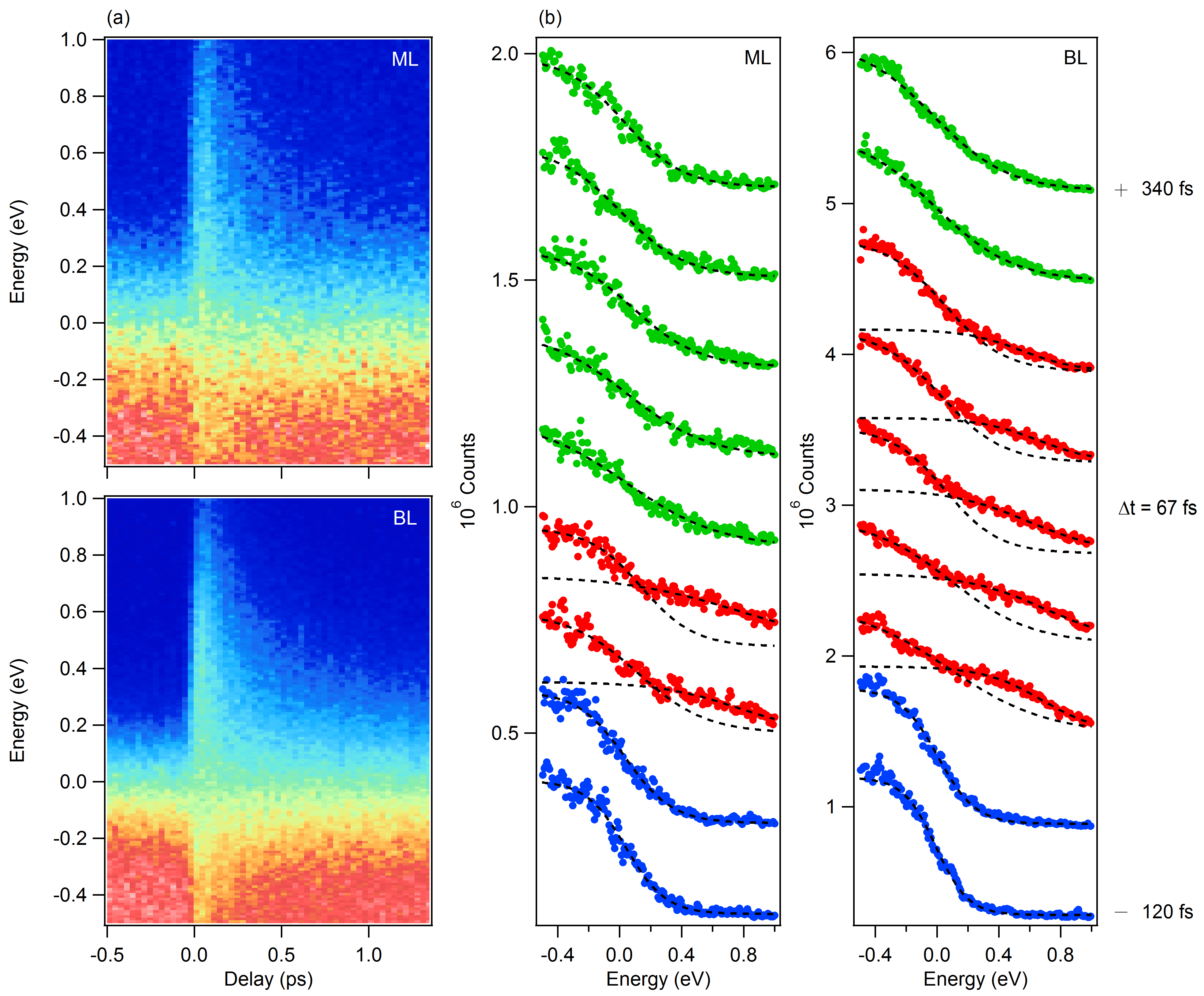}
  \caption{(a) Momentum-integrated photocurrent for monolayer and bilayer graphene as a function of pump-probe delay close to the Fermi level. (b) Momentum-integrated energy-distribution curves for selected pump-probe delays around time zero (data points) together with Fermi-Dirac fits (black dashed lines) for monolayer and bilayer graphene. Different colors indicate different regimes corresponding to the formation (change from blue to red) and reunification (change from red to green) of decoupled electron gases in valence and conduction band. Time-steps are indicated on the right side of the figure.}
  \label{fig_FD}
\end{figure}

A more detailed data analysis is presented in Fig. \ref{fig_FD}. In Fig. \ref{fig_FD}a we plot the momentum-integrated photocurrent as a function of pump-probe delay for monolayer and bilayer graphene. Figure \ref{fig_FD}b displays momentum-integrated energy distribution curves for selected pump-probe delays around time zero, where the pump-probe signal is maximum. The dashed black lines represent fits to the experimental data using Fermi-Dirac (FD) distribution functions. At negative delays (blue data points), the data can be nicely described by a single FD distribution. The shape of the non-equilibrium distributions around time zero (red data points), however, clearly deviates from a single FD distribution. Instead, we use two separate FD distributions, one for the valence band and one for the conduction band, to fit the data. This indicates that, at earliest times after interband excitation, a relaxation bottleneck at the Dirac point causes population inversion. The two electron gases rapidly thermalize into separate FD distributions with different temperatures and chemical potentials. The two FD distributions merge (green data points) after $\sim130$\,fs and $\sim330$\,fs for monolayer and bilayer samples, respectively. A single FD distribution at elevated electronic temperature is then observed.

\begin{figure}
	\center
  \includegraphics[width = 1\columnwidth]{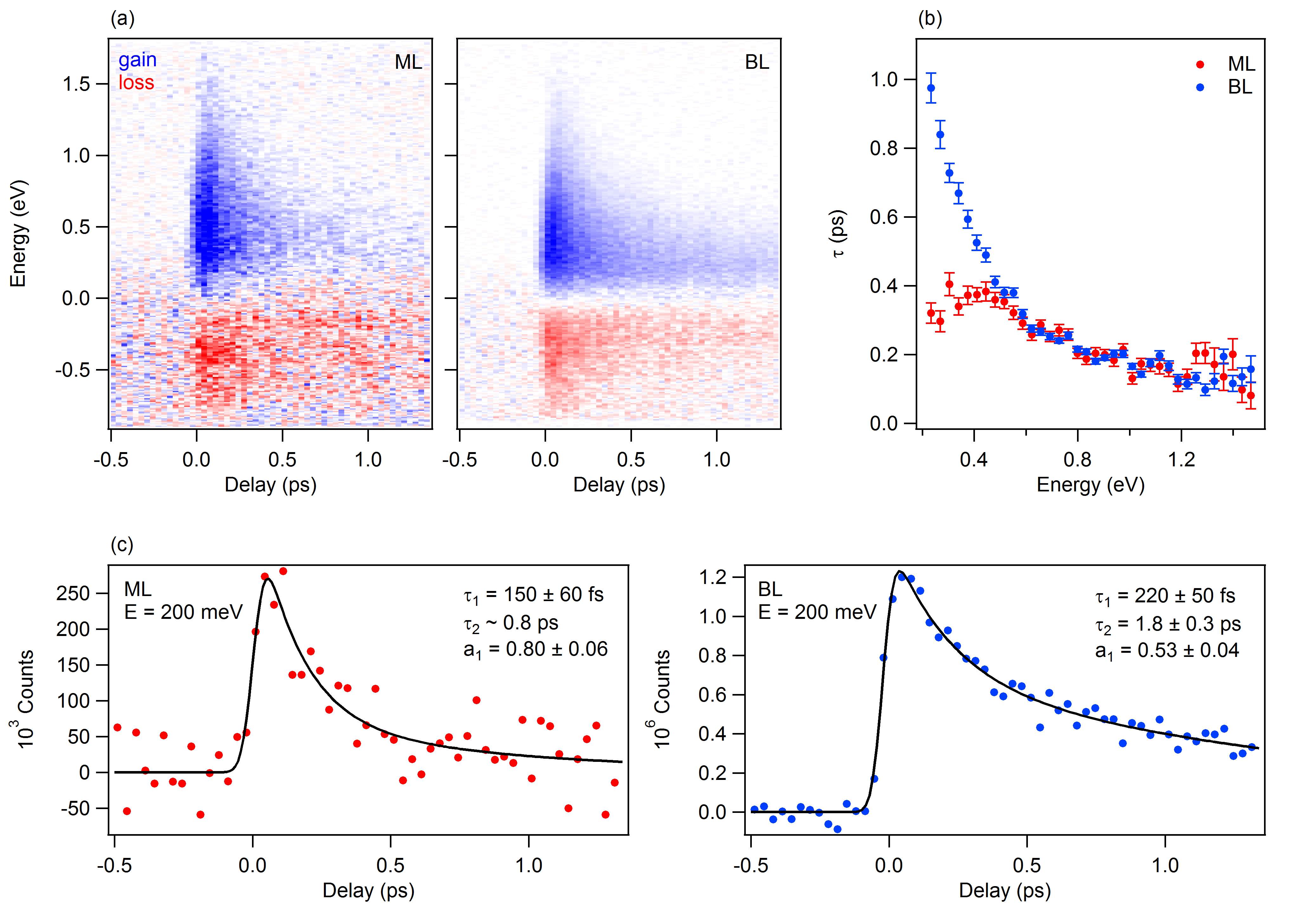}
  \caption{(a) Momentum-integrated pump-induced changes of the photocurrent for monolayer and bilayer graphene. (b) Exponential decay time as a function of energy for monolayer (red) and bilayer graphene (blue). (c) Line profiles at constant energy $E=200$\,meV taken from panel a (data points) together with double exponential fits (continuous black lines) for monolayer (red) and bilayer graphene (blue).}
  \label{fig_tau}
\end{figure}

In Fig. \ref{fig_tau} we analyze the energy dependence of the relaxation time. In Fig. \ref{fig_tau}a we plot the momentum-integrated pump-induced changes of the photocurrent as a function of pump-probe delay for monolayer and bilayer graphene. From these plots we extract line profiles at constant energy and fit them with an exponentially decaying function. The resulting relaxation times as a function of energy are plotted in Fig. \ref{fig_tau}b. In both monolayer and bilayer graphene the relaxation rates increase with decreasing energy with a long-lived pump-probe signal close to the Fermi level. For energies above $E=0.5$\,eV the decay times for monolayer and bilayer graphene are comparable. For energies smaller than $E=0.5$\,eV, however, charge carriers in bilayer graphene relax much more slowly than in monolayer graphene. Note that, for small energies, the relaxation dynamics is better described by a double exponential decay. One example for this is shown in Fig. \ref{fig_tau}c, where line profiles at $E=200$\,meV are extracted from Fig. \ref{fig_tau}a and plotted together with double exponential fits (continuous black lines). Also here, the decay times in the bilayer ($\tau_1=220\pm50$\,fs, $\tau_2=1.8\pm0.3$\,ps) are longer than the ones in the monolayer ($\tau_1=150\pm60$\,fs, $\tau_2\sim0.8$\,ps). Furthermore, the amplitude of the fast component, $a_1$, is larger in monolayer ($a_1=0.80\pm0.06$) than in bilayer graphene ($a_1=0.53\pm0.04$). This is in agreement with the intuitive assumption that the actual band gap of $\sim200$\,meV in bilayer graphene imposes a bigger obstacle to carrier relaxation than the zero density of states at the Dirac point in monolayer graphene. 

\section{Discussion}

\subsection{Relaxation Dynamics}

Interband transitions at $\hbar\omega_{\text{pump}}$ generate holes in the valence band at $E=E_D-\hbar\omega_{\text{pump}}/2$ and electrons in the conduction band at $E=E_D+\hbar\omega_{\text{pump}}/2$, where $E_D$ is the energy of the Dirac point where valence and conduction band touch. The initial carrier distribution is highly anisotropic with nodes along the direction of pump polarization and maxima in the direction perpendicular to the pump polarization \cite{Malic_2011,Mittendorff_2014}. Collinear electron-electron scattering results in an ultrafast thermalization of the non-equilibrium distribution within $\sim10$\,fs with a momentum-dependent electronic temperature  \cite{Butscher_2007,George_2008,Sun_2008,Breusing_2009,Sun_2010_2,Lui_2010,Breusing_2011,Malic_2011}. This distribution becomes isotropic and cools down by the emission of optical phonons on a timescale of $\sim200$\,fs, establishing a quasi-equilibrium between electrons and hot optical phonons \cite{Kampfrath_2005,Bonini_2007,Yan_2009,Wang_2010,Kang_2010,Huang_2010,Hale_2011,Chatzakis_2011}. This quasi-equilibrium further cools down by the decay of optical into acoustic phonons \cite{Sun_2008,Lui_2010,Kampfrath_2005,Bonini_2007,Yan_2009,Wang_2010,Kang_2010,Huang_2010,Zou_2010,Hale_2011,Chatzakis_2011,Winnerl_2011} or a direct coupling of the electrons to acoustic phonons via supercollisions in the presence of defects \cite{Song_2012,Graham_2013,Betz_2013,Johannsen_2013} within a few picoseconds.

The earliest stages of the relaxation have not been analyzed extensively to date due to the lack of the required time and/or momentum resolution. Breusing \etal \cite{Breusing_2009,Breusing_2011}, with a temporal resolution of $\sim10$\,fs, determined an initial ultrafast thermalization time due to electron-electron scattering of $\tau_0=13\pm3$\,fs. They also pointed out that two separate Fermi-Dirac distributions for valence and conduction band are required to properly fit their differential transmission data. This interpretation has been supported by first angle-integrated time-resolved photoemission (tr-PES) experiments on multidomain graphene samples by Gilbertson \etal \cite{Gilbertson_2011,Gilbertson_2012}. Mittendorff \etal \cite{Mittendorff_2014} have shown recently that the initial distribution of photoexcited carriers stays anisotropic for about $\sim$150\,fs after excitation before optical phonon emission results in efficient angle relaxation. 

The rapid thermalization of photoexcited carriers is in good agreement with our data in Fig. \ref{fig_FD}, where the carriers follow a thermal FD distribution at all times. Further, our measurements give direct evidence for the existence of independent electron gases in the valence and conduction band at early times, supporting the interpretation of previous optical pump-probe \cite{Breusing_2009,Breusing_2011} and tr-PES experiments \cite{Gilbertson_2011,Gilbertson_2012,Gierz_2013}.

Fast cooling of the hot electronic distribution via the emission of strongly coupled optical phonons within $\tau_1\sim200$\,fs has been demonstrated unequivocally using time-resolved anti-Stokes Raman scattering, directly probing the in-plane $E_{2g}$ mode occupancy as a function of time \cite{Yan_2009,Kang_2010,Chatzakis_2011}. These measurements also suggest that the slow relaxation time $\tau_2$ of several picoseconds can be associated with the anharmonic decay of hot optical phonons into acoustic phonons. Recently, it has been pointed out that $\tau_2$ might also be due to supercollisions, where disorder-assisted scattering events enable a more efficient heat transfer from the electronic system directly to acoustic phonons \cite{Song_2012,Graham_2013,Betz_2013,Johannsen_2013}.

\subsection{Relaxation Bottleneck and Population Inversion}

The relaxation bottleneck in graphene has two different contributions. The first contribution is the zero density of states (ZDOS) at the Dirac point in monolayer graphene. This \emph{ZDOS bottleneck} reduces the phase space for any electronic transition involving states close to the Dirac point, independent of whether these transitions are mediated by electron-electron or electron-phonon scattering. The second contribution is related to the fact that the main cooling channel for hot electrons is the emission of strongly coupled optical phonons. Electrons in graphene mainly couple to $E_{2g}$ phonons with momentum $q\approx\Gamma$ and energy $E=195$\,meV, and $A_1'$ phonons with $q\approx K$ and $E=160$\,meV \cite{Calandra_2007,Park_2008}. Thus, hot electrons can only cool down via optical phonon emission, if their excess energy with respect to the equilibrium electron distribution is higher than the phonon energy. For excess energies smaller than the phonon energy, phonon emission is not possible, resulting in very long relaxation times \cite{Winnerl_2011,Sentef_2013}. This contribution is referred to as the \emph{phonon bottleneck}. We would like to point out that the ZDOS bottleneck is responsible for an accumulation of carriers at the bottom of the conduction band, while the phonon bottleneck accounts for long lifetimes in the vicinity of the Fermi level.

In 2007, Ryzhii \etal \cite{Rhyzhii_2007} predicted that strong optical excitation may result in population inversion in graphene with electrons at the bottom of the conduction band and holes at the top of the valence band. The effect was attributed to the ZDOS bottleneck and is expected to result in negative optical conductivities in the Terahertz regime. The relaxation bottleneck in graphene was found to be strong enough to allow for saturable absorption, which was initially attributed to the formation of a hot Fermi-Dirac distribution that results in Pauli blocking of the optical transition \cite{Bao_2009,Sun_2010}. Recently, Li \etal \cite{Li_2012} measured a transient negative optical conductivity at 1.16\,eV probe energy for pump fluences above 2\,mJ/cm$^2$ at $\hbar\omega_{\text{pump}}=1.55$\,eV. These observations have been confirmed with time- and angle-resolved photoemission spectroscopy \cite{Gilbertson_2011,Gilbertson_2012,Gierz_2013} and explained by microscopic calculations \cite{Winzer_2013}. According to Winzer \etal \cite{Winzer_2013} population inversion after optical excitation arises due to rapid phonon-induced intraband scattering that fills the states at the bottom of the conduction band, where the ZDOS bottleneck restricts further relaxation. The inverted population then decays within a few hundred femtoseconds due to Auger recombination.

\begin{figure}
	\center
  \includegraphics[width = 0.7\columnwidth]{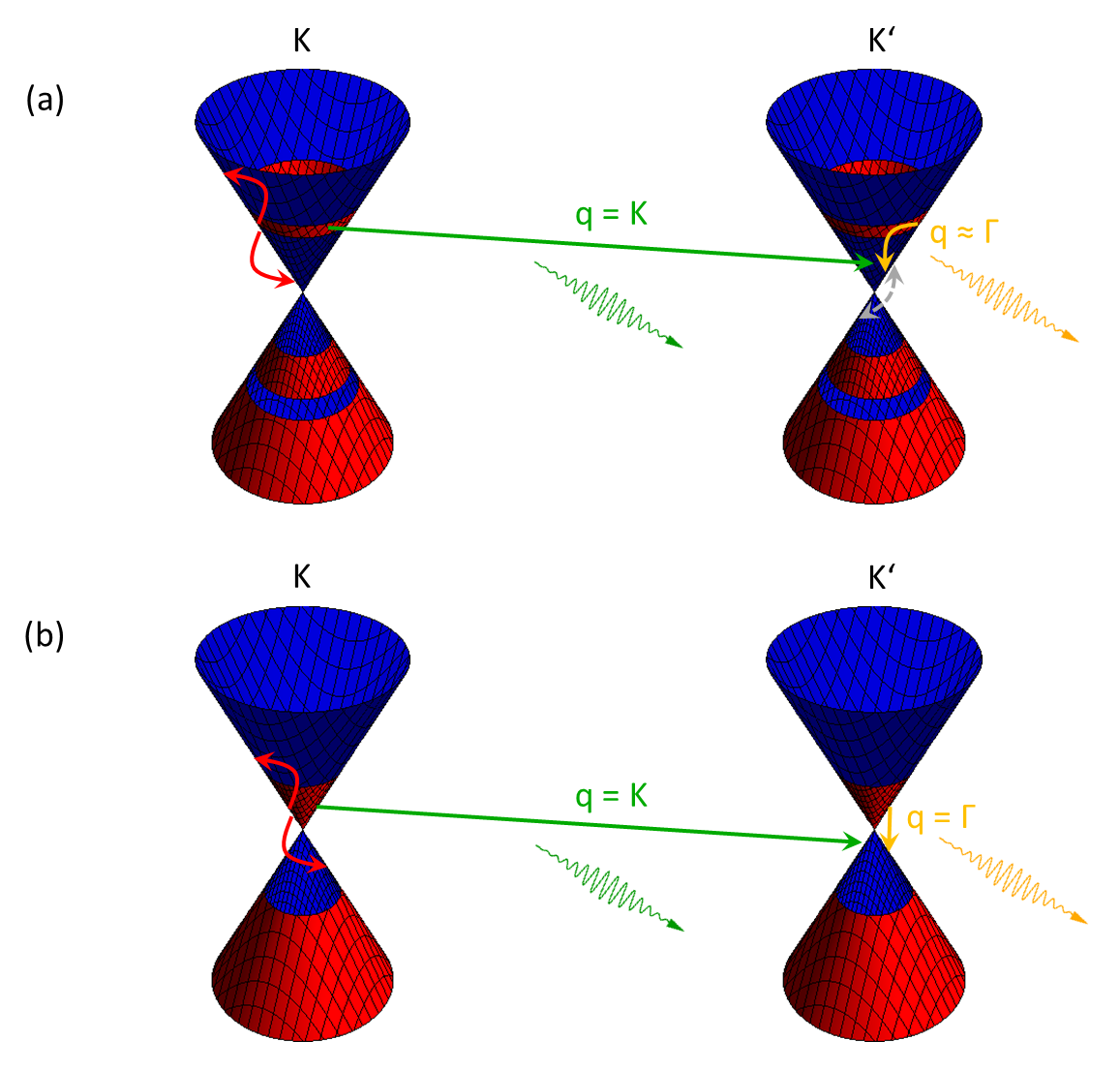}
  \caption{Electron-electron and electron-phonon scattering channels in photo-excited graphene. Direct interband transitions generate holes in the valence band and electrons in the conduction band. Occupied states are shown in red, empty states are shown in blue. During and immediately after excitation (a) the excited electrons rapidly thermalize via Auger scattering (red arrows) and cool down via the emission of K-point phonons with $q=K$ (green arrow). Emission of phonons with $q\approx\Gamma$ (yellow arrow) leads to the accumulation of carriers at the bottom of the conduction band. Supported by the suppression of interband scattering due to the slight hole-doping of our samples (dashed grey arrow) the ZDOS bottleneck results in a short-lived population inversion with independent electron gases in valence and conduction band. Once an inverted population is established (b) efficient interband scattering dominated by fast Auger recombination rapidly merges the separate chemical potentials and results in a single Fermi-Dirac distribution for both valence and conduction band. The hot electron gas cools down via the emission of optical phonons at $\Gamma$ and K.}
  \label{fig_scattering}
\end{figure}

In Fig. \ref{fig_scattering} we present the microscopic picture introduced by Winzer \etal \cite{Winzer_2013}. This picture accounts for the occurrence of population inversion in graphene (Fig. \ref{fig_scattering}a) and its limited lifetime (Fig. \ref{fig_scattering}b). Photoexcitation transfers electrons from $E=E_D-\hbar\omega_{\text{pump}}/2$ to $E=E_D+\hbar\omega_{\text{pump}}/2$. Due to the absence of charge carriers in the vicinity of the Dirac point in our hole-doped samples, interband scattering is inhibited, efficiently decoupling valence and conduction band. Auger scattering rapidly thermalizes the independent electron gases in the two bands, while the emission of $E_{2g}$ and $A_1'$ optical phonons leads to an accumulation of charge carriers at the bottom of the conduction band (see Fig. \ref{fig_scattering}a). Once the inverted population is established (see Fig. \ref{fig_scattering}b), Auger recombination sets in, restoring one common FD distribution for valence and conduction band. From this point on, the hot electrons cool down via optical phonon emission until a quasi-equilibrium between hot electrons and hot optical phonons is reached. This quasi-equilibrium further cools by the emission of acoustic phonons.

The actual band gap in bilayer graphene restricts the available phase space for Auger recombination compared to monolayer graphene and thus extends the life time of the inverted population from $\sim130$\,fs to $\sim330$\,fs.

\subsection{Comparison with Previous tr-ARPES Data on Graphene}

Only few time-resolved photoemission investigations of charge carrier dynamics in graphene exist. Not all of them allow for a direct comparison with our findings, as in some cases the dynamics were mainly dominated by the influence of the substrate. 

Gilbertson \etal \cite{Gilbertson_2011,Gilbertson_2012} investigated graphene grown by chemical vapor deposition on copper and transferred to SiO$_2$. Although their samples only allowed for an angle-integrated investigation they observed a transition from two separate into a single FD distribution during the relaxation process for an incident pump fluence of $F=1.6$\,mJ/cm$^2$ at $\hbar\omega_{\text{pump}}=1.55$\,eV \cite{Gilbertson_2011}. They attributed the observed energy dependence of the quasiparticle lifetime to the ZDOS bottleneck \cite{Gilbertson_2012} although --- considering Ref \cite{Sentef_2013} and the above discussion --- it is probably rather due to the phonon bottleneck, as the maximum lifetime coincides with the Fermi level and not the Dirac point.

For graphene on metal substrates such as Ru(0001) \cite{Armbrust_2012} and Ni(111) \cite{Bignardi_2014} the lifetimes of the excited carriers in the vicinity of the Fermi level are dominated by the substrate, resulting in much faster decay times on the order of a few tens of femtoseconds that are not related to intrinsic graphene dynamics.

Finally, Johannsen \etal \cite{Johannsen_2013} and Ulstrup \etal \cite{Ulstrup_2014} performed very similar experiments compared to our own also using quasi-freestanding graphene monolayer and bilayer samples on SiC(0001), albeit at lower pump fluences. In Ref. \cite{Johannsen_2013}, for a pump fluence of 0.3\,mJ/cm$^2$, the authors observe a single FD distribution at all times. The data analysis focused on the cooling dynamics of the Dirac carriers, revealing that supercollisions play an important role in energy dissipation. The authors were also looking for charge carrier multiplication \cite{Winzer_2010,Winzer_2012,Brida_2012}, but concluded that the required low-fluence regime is not accessible with the 1\,kHz experimental setup employed, in agreement with our own findings \cite{Gierz_2013}. Ref. \cite{Ulstrup_2014} aims at a detailed comparison of the different dynamics observed in monolayer and bilayer graphene. In good agreement with our own data in Fig. \ref{fig_tau}b and c, the authors report pronounced differences in the dynamics only at low energies, consistent with a stronger relaxation bottleneck in the presence of a band gap. 

However, Ulstrup \etal \cite{Ulstrup_2014} find that, even at a comparatively high pump fluence of 3.5\,mJ/cm$^2$ at $\hbar\omega_{\text{pump}}=1.55$\,eV, the pump-probe data on bilayer graphene can be described with a single FD distribution broadened by the experimental resolution. Their momentum-integrated data at the peak of the response exhibits a clear shoulder in the conduction band that they attribute to an accumulation of carriers in the upper branch of the conduction band due to the elevated electronic temperature. At first glance this may seem to contradict our interpretation of a similar shoulder (see Fig. \ref{fig_FD}b) in terms of a second separate FD distribution. However, the shoulders reported here (Fig. \ref{fig_FD}b) are much stronger than the one observed by Ulstrup \etal \cite{Ulstrup_2014}. Also, we observe a similar shoulder in monolayer graphene with only one set of $\pi$-bands, clearly indicating that the interpretation given by Ulstrup \etal does not apply in our case. 

Finally, the amplitude of the second higher-energy FD distribution in Fig. \ref{fig_FD}b is lower than the one for the low-energy FD distribution, which is not expected for an electron-hole symmetric density of states. We believe that this apparent discrepancy may come from the fact that the two separate FD distributions are only observed at early times where the hot carriers are still anisotropically distributed in momentum space \cite{Malic_2011,Mittendorff_2014}. In the present study, the pump polarization lies along the $\Gamma$K-direction. Thus, the ARPES cuts are taken along the nodes of the initially anisotropic carrier distribution, and photoexcited electrons will only appear in the field of view of the ARPES cut after phonon-mediated momentum relaxation has taken place, causing the apparent reduction of density of states in the conduction band.

\section{Conclusion}

In summary, we have used time- and angle-resolved EUV photoemission to investigate hot-electron relaxation in monolayer and bilayer graphene. In the strong excitation regime used here, we find that the lifetime of the inverted population is longer in bilayer than in monolayer graphene. This is interpreted as a direct consequence of the different electronic structures. These results indicate that the relaxation bottleneck associated with the zero density of states at the Dirac point in monolayer graphene is enhanced by the ~200 meV band gap found in bilayer graphene. The observation of population inversion may allow for applications of graphene as an amplifying medium for Terahertz radiation. For bilayer graphene, the small band gap limits the operation regime of such an amplifier to frequencies above 50\,THz (6\,$\mu$m wavelength). Finally, further enhancement of the lifetime of the inverted population may be possible at higher pump energies in combination with higher pump fluences, eventually enabling THz lasing.

\section{Acknowledgments}

The research leading to these results has received funding from LASERLAB-EUROPE (grant agreement no. 284464, EC's Seventh Framework Programme).

\end{document}